\documentclass[sigconf,screen]{acmart}
\setcopyright{rightsretained}
\acmPrice{}
\acmDOI{10.1145/3338906.3340458}
\acmYear{2019}
\copyrightyear{2019}
\acmISBN{978-1-4503-5572-8/19/08}
\acmConference[ESEC/FSE '19]{Proceedings of the 27th ACM Joint European Software Engineering Conference and Symposium on the Foundations of Software Engineering}{August 26--30, 2019}{Tallinn, Estonia}
\acmBooktitle{Proceedings of the 27th ACM Joint European Software Engineering Conference and Symposium on the Foundations of Software Engineering (ESEC/FSE '19), August 26--30, 2019, Tallinn, Estonia}

\usepackage{amsmath,amssymb,amsfonts}
\usepackage{algorithmic}
\usepackage{graphicx}
\usepackage{textcomp}
\usepackage{xcolor}
\usepackage{hyperref}
\usepackage{cleveref}
\usepackage{booktabs}
\usepackage{subcaption}
\usepackage{listings}
\usepackage{courier}
\usepackage{array}
\usepackage{verbatim}
\usepackage{caption}
\usepackage{cancel}
\usepackage{multirow}
\usepackage[subtle]{savetrees}

\usepackage{stmaryrd}
\usepackage{enumitem}
\usepackage{balance}

\DeclareCaptionFont{white}{\color{white}}
\DeclareCaptionFormat{listing}{
  \colorbox[cmyk]{0.1, 0.1, 0.1, 1}{
    \parbox{\dimexpr\linewidth-3\fboxsep}{#1#2#3}
  }
}
\definecolor{javared}{rgb}{0.75, 0, 0}
\definecolor{javapurple}{rgb}{0.47, 0.36, 0.64}
\definecolor{javablue}{rgb}{0, 0.36, 0.77}

\def\BibTeX{{\rm B\kern-.05em{\sc i\kern-.025em b}\kern-.08em
    T\kern-.1667em\lower.7ex\hbox{E}\kern-.125emX}}
\begin{document}

\title{When Deep Learning Met Code Search}

\newcommand{\todo}[1]{{\textcolor{red}{\textbf{TODO}: #1}}}
\newcommand{\ksen}[1]{{\textcolor{blue}{\textbf{KS}: #1}}}
\newcommand{\jose}[1]{{\textcolor{blue}{\textbf{JOSE}: #1}}}
\newcommand{\hongyu}[1]{{\textcolor{javablue}{\textbf{HONGYU}: #1}}}
\newcommand{\sonia}[1]{{\textcolor{violet}{\textbf{SONIA}: #1}}}

\newcommand{\codenn}[0]{\textit{CODEnn}}
\newcommand{\cooc}[0]{\textit{COOC}}
\newcommand{\eu}[0]{\textit{UNIF}}
\newcommand{\githubmodel}[0]{\textit{SCS}}
\newcommand{\ncs}[0]{\textit{NCS}}

\newcommand{\dcstrain}[0]{\textit{\codenn{}-Java-Train}}
\newcommand{\androidtrain}[0]{\textit{GitHub-Android-Train}}
\newcommand{\sotrain}[0]{\textit{StackOverflow-Android-Train}}

\newcommand{\dcssearch}[0]{\textit{\codenn{}-Java-Search}}
\newcommand{\androidsearch}[0]{\textit{GitHub-Android-Search}}

\newcommand{\ncsquestions}[0]{\textit{Android-287}}
\newcommand{\dcsquestions}[0]{\textit{Java-50}}

\newcommand{\debugfixed}[0]{\textit{\eu{}-F}}
\newcommand{\debugsupervised}[0]{\textit{\eu{}-S}}


\author{Jos\'e Cambronero}
\authornote{Both authors contributed equally to the paper}
\affiliation{
  \institution{MIT CSAIL}
  \country{U.S.A.}
}
\email{jcamsan@mit.edu}

\author{Hongyu Li}
\authornotemark[1]
\affiliation{
  \institution{Facebook, Inc.}
  \country{U.S.A.}
}
\email{hongyul@fb.com}

\author{Seohyun Kim}
\affiliation{
  \institution{Facebook, Inc.}
  \country{U.S.A.}
}
\email{skim131@fb.com}

\author{Koushik Sen}
\affiliation{
  \institution{EECS Department, UC Berkeley}
  \country{U.S.A.}
}
\email{ksen@cs.berkeley.edu}

\author{Satish Chandra}
\affiliation{
  \institution{Facebook, Inc.}
  \country{U.S.A.}
}
\email{schandra@acm.org}

\begin{abstract}
There have been multiple recent proposals on using deep
neural networks for code search using natural language.
Common across these proposals is the idea of \textit{embedding}
code and natural language queries into real vectors and then
using vector distance to approximate semantic correlation between code and
the query. Multiple approaches exist for learning these embeddings,
including \emph{unsupervised} techniques, which rely only on a corpus of code
examples, and \emph{supervised} techniques, which use an \emph{aligned}
corpus of paired code and natural language descriptions. The goal of this supervision is to produce
embeddings that are more similar for a query and the corresponding desired code snippet.

Clearly, there are choices in whether to use supervised techniques at all,
and if one does, what sort of network and training to use for supervision.
This paper is the first to evaluate these
choices systematically.  To this end, we assembled implementations of
state-of-the-art techniques to run on a common platform, training
and evaluation corpora. To explore the
design space in network complexity, we also introduced a new design point
that is a \emph{minimal} supervision extension to an existing unsupervised technique.

Our evaluation shows that: 1. adding supervision to an existing unsupervised
technique can improve performance, though not necessarily by much; 2. simple
networks for supervision can be more effective that more
sophisticated sequence-based networks for code search; 3. while it is common
to use docstrings to carry out supervision, there is a sizable gap
between the effectiveness of docstrings and a more query-appropriate supervision
corpus.
\end{abstract}

\ccsdesc[500]{Computing methodologies~Learning latent representations}
\ccsdesc[500]{Software and its engineering~General programming languages}
\ccsdesc[500]{Software and its engineering~Reusability}

\keywords{code search, neural networks, joint embedding}

\maketitle

\section{Introduction}
We have recently seen a significant uptick in interest in code search.
The goal of code search is to retrieve code fragments from a large code
corpus that most closely match a developer's intent, which is expressed
in natural language.
Being able to examine existing code that is relevant to a developer's
intent is a fundamental productivity tool. Sites such as Stack Overflow are popular
because they are easy to search for code relevant to a user's
question expressed in natural language.


Proprietary code repositories in particular pose a challenge, as developers
can no longer rely on public sources such as Google or Stack Overflow
for assistance, as these may not capture the required organization-specific
API and library usage. However,
recent works from both academia
and industry~\cite{deepcodesearch, sachdev2018retrieval, githubblog, iyer2016summarizing} have taken steps towards enabling more advanced code search using deep learning.
We call such methods \textit{neural} code search.  See
Figure~\ref{tab:examples} for some examples of code snippets retrieved based on natural
language queries: it is evident that the state of the technology has
become promising indeed.  The type of queries presented in Figure~\ref{tab:examples}, and
the accompanying results, also highlight the difficulties associated with
tackling this task purely based on simple approaches such as regular-expression
matching.

\begin{figure}

\lstset{
    language=Java,
    basicstyle=\scriptsize\ttfamily, 
    tabsize=2,                       
    extendedchars=true,
    breaklines=true,                 
    keywordstyle=\color{javared}\bf\ttfamily,
    frame=single,
    stringstyle=\color{javablue}\bf\ttfamily, 
    commentstyle=\color{javapurple},
    showspaces=false,
    showtabs=false,
    xleftmargin=6.3pt,
    framexleftmargin=2.05pt,
    framexrightmargin=-2.95pt,
    framexbottommargin=4pt,
    framextopmargin=4pt,
    showstringspaces=false,
    escapeinside={<@}{@>}
}

\begin{lstlisting}[
  label={lst:query1},
  caption={How can I convert a stack trace to a string?}
]
public synchronized static String <@\textbf{\textcolor{javapurple}{getStackTrace}}@>(Exception e) {
  e.fillInStackTrace();
  StringBuffer buffer = new StringBuffer();
  buffer.append(e.getMessage() + "-");
  for (StackTraceElement el: e.getStackTrace()) {
    buffer.append(el.toString() + "-");
  }
  return buffer.toString();
}
\end{lstlisting}
\begin{flushleft}
\tiny\url{https://github.com/Dynatrace/Dynatrace-AppMon-REST-Monitor-Plugin/blob/master/src/com/realdolmen/dynatrace/restmonitor/RestMonitor.java}
\end{flushleft}

\hrulefill

\begin{lstlisting}[
  label={lst:query-2},
  caption={How do I get a platform-dependent new line character?}
]
public static String <@\textbf{\textcolor{javapurple}{getPlatformLineSeparator}}@>() {
  return System.getProperty("line.separator");
}
\end{lstlisting}
\begin{flushleft}
\tiny\url{https://github.com/nutritionfactsorg/daily-dozen-android/blob/master/app/src/main/java/org/nutritionfacts/dailydozen/Common.java}
\end{flushleft}

\hrulefill

\caption[caption for code examples]{Example code search results.  Each is selected from the top 1 result found by the \eu{} model that we introduce. The existing code search interface of github.com does not return any relevant code snippets in the top 10 results for these queries.}
\label{tab:examples}
\end{figure}

\begin{figure*}[!h]
  \centering
  \includegraphics[width=\textwidth]{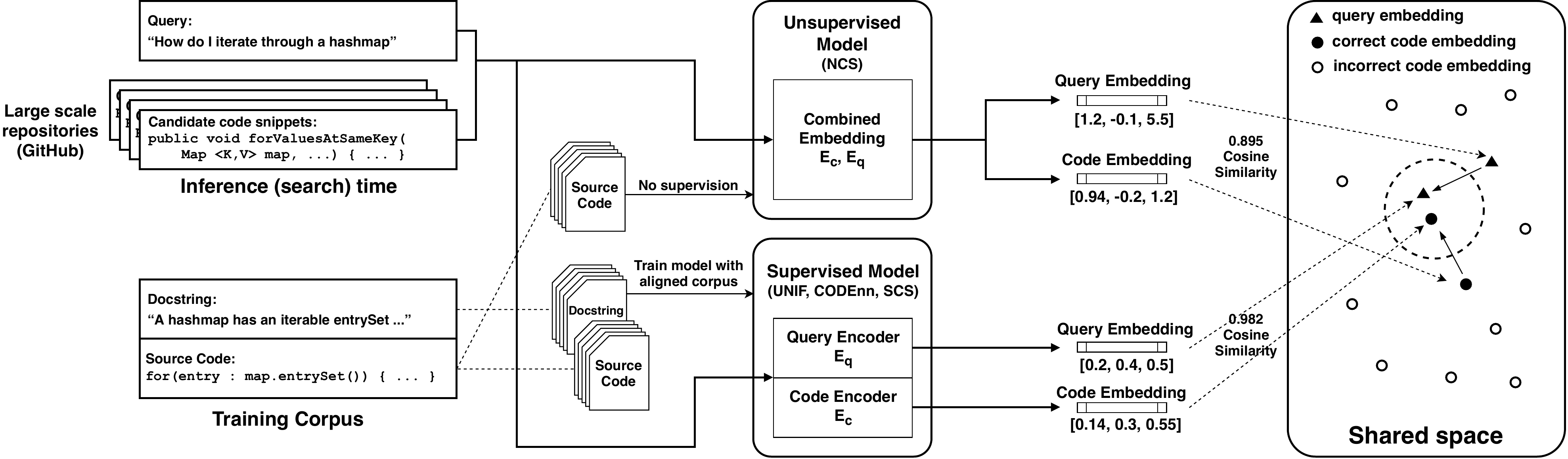}
  \caption{When using embeddings for code search, the query and the
  candidate code snippets are mapped to a shared vector space,
  using functions $E_q$ and $E_c$, respectively, and maximizing cosine similarity between corresponding query embedding and code embeddings.
  These vector representations can be learned in an \emph{unsupervised} manner, which just uses code,
  or in a \emph{supervised} manner, which exploit an \emph{aligned} corpus of code and their
  corresponding natural language descriptions.
  }
  \label{fig:example}
\end{figure*}

\Cref{fig:example} provides a general overview of neural code search and
outlines different techniques, which we address in detail
through this paper.
The core abstraction in neural code search systems
is the notion of \emph{embeddings}, which are vector
representations of inputs in a shared vector space.
By computing a vector similarity measure, such as cosine similarity~\cite{nncosine}, over these embeddings,
search can retrieve code fragments that are semantically related to a user query.
For example, in \Cref{fig:example}, the query \textit{``How do I iterate through
a hashmap?''} is mapped to the vector $\langle 1.2, -0.1, 5.5 \rangle$
by one possible technique (\ncs{}).
Candidate code snippets are also mapped to vectors using the same technique. In \Cref{fig:example},
one such code snippet, \texttt{public void forValuesAtSameKey...}, for example, is mapped
to the vector $\langle 0.94, -0.2, 1.2 \rangle$.
The candidate code snippets then can be ranked using vector similarity.
A key challenge in neural code search is to
learn these embeddings in a way that vector similarity
coincides with semantic relatedness.

As shown in \Cref{fig:example}, the models used to learn these representations
can be broadly grouped into \emph{unsupervised} and \emph{supervised}.
In our journey to explore the advantages of neural techniques, we started with \ncs{}, an effective, unsupervised, neural code search technique we previously built in~\cite{sachdev2018retrieval}. Because \ncs{} showed promising results, we wanted to experiment
with the possibility of improving upon this baseline through additional design
enhancements. In particular, recent work~\cite{deepcodesearch, githubblog}
presented promising supervised neural code search techniques, labeled \codenn{} and \githubmodel{} respectively, that successfully
learned code and natural language embeddings using corpora of source code and
docstrings.

The goal of this supervision is to learn a mapping that produces
more similar vectors for user queries
and the corresponding desired code. In \Cref{fig:example}, this goal is
depicted by the solid arrows, which move the embeddings for the query
and correct code fragment closer together when mapped using a supervised model.

With so many techniques to choose from, how does anyone trying to design and deploy a
code search solution make an informed choice?
For instance, supervision sounds like a good idea, but how much benefit does it provide,
relative to the overhead of obtaining the supervision data? How much value, if any, do
the more sophisticated networks -- which have many more parameters -- bring compared to a
simpler network, one of which we introduce in this work (\eu{} in \Cref{fig:example},
described further below)?
Does model supervision carried out using docstrings
potentially limit performance when models are applied to real user queries?

In this work, we attempt to understand these tradeoffs quantitatively.
To do so, we formulate experiments in the context of the code search
techniques mentioned above.
Three of these techniques are exactly as in previous work, and are
state-of-the-art at this time:
\begin{itemize}[leftmargin=*]
\item  \textbf{\ncs}: An unsupervised technique for neural code search developed at
 Facebook~\cite{sachdev2018retrieval}, which uses only word embeddings derived
 from a code corpus.
 %

\item \textbf{\codenn}: A supervised technique from a recent paper on code search using deep
neural networks~\cite{deepcodesearch}, which uses multiple
sequence-to-sequence-based networks,
and was shown to outperform other state-of-the-art code search techniques.
We use the implementation provided by the authors~\cite{dcsgithub}.

\item \textbf{\githubmodel}: A supervised neural code search system using multiple sequence-to-sequence networks.
We use the implementation provided by the authors in a blog post~\cite{githubblog, githubblog2}.
\end{itemize}

\noindent
Because we wanted to understand the extent to which the complex sequence-of-words based
networks help, we also developed a minimal extension to the \ncs{} technique, just adding
supervision and nothing else:

\begin{itemize}[leftmargin=*]
\item \textbf{\eu{}}: A supervised extension of the base
\ncs{} technique of our own creation. \eu{} uses a bag-of-words-based network, which has
significantly lower complexity compared to sequence-of-words-based networks.
This simple model is a new contribution of this paper.
\end{itemize}

Our evaluation is structured using the following three research questions;
we also give the summary of our findings along with the question.

\paragraph*{Research Question 1}: Does extending an effective
  unsupervised code search technique with supervision based on
  a corpus of paired code and natural language descriptions improve performance?

  Our results show that \eu{} performed better than \ncs{}
  on our benchmark queries, though the improvement
  was not seen uniformly across all data sets.


\paragraph*{Research Question 2}: Do more sophisticated networks
  improve performance in supervised neural code search?

  Our results show that \eu{}, a simple, bag-of-words-based network, outperformed
  the sequence-of-words-based \codenn{} and
  \githubmodel{} on our benchmarks.
  The additional sophistication did not add value.


\paragraph*{Research Question 3}: How effective is supervision based on docstrings
as the natural language  component of the training corpus?

We found that supervision based on docstrings -- which is commonly the natural
language component of an aligned corpus -- did not always improve
performance, contrary to expectations. To understand the possible
room for improvement, we constructed an ideal alternate training corpus, where
the code snippets and natural language, while disjoint from our
benchmark queries, were drawn from the same source.

When trained on this corpus, all supervised techniques improved significantly,
showing that, as a proof of concept, if given a training corpus that
matches expected user evaluation, these techniques can provide
impressive search performance.




\paragraph*{\bf Contributions}
\noindent
1.  We believe this is the first paper to compare recent neural code search
systems running on the same platform and evaluation using the same corpora.

\noindent
2. We present a new design point in the spectrum of neural code search
systems: \eu{}, an extension to \ncs{} that minimally adds supervision and
nothing else.

\noindent
3. Our findings are that \eu{} outperforms some of the more sophisticated
network designs (\codenn{} and \githubmodel{}) as well as \ncs{}, the unsupervised technique. Moreover,
the choice of the aligned corpus used in supervision is extremely pertinent:
an idealized training corpus shows that supervised techniques can deliver
impressive performance, and highlights the differences in performance
that may not be immediately evident from training on the typical code and docstring aligned
corpora.


These findings have implications for anyone considering designing and
deploying a neural code search system in their organization.  The
findings also have implications for researchers, who should consider
simple baseline models in evaluating their designs.

\paragraph*{\bf Outline} The rest of the paper is organized as follows.
\Cref{sec:emb} introduces the core idea of embeddings and their use
in neural code search. \Cref{sec:models} details each of the techniques
explored in this paper. \Cref{sec:methodology} presents our evaluation
methodology. \Cref{sec:results} provides results supporting
our research questions' answers.
\Cref{sec:threats} and \Cref{sec:relatedwork} discuss threats to validity
and related work, respectively. Finally,
\Cref{sec:conclusion} concludes with the main takeaways and implications
for neural code search system designers.

\lstset{
    language=Java,
    basicstyle=\footnotesize\ttfamily, 
    tabsize=2,                       
    extendedchars=true,
    breaklines=true,                 
    keywordstyle=\color{javared}\bf\ttfamily,
    frame=single,
    stringstyle=\color{javablue}\bf\ttfamily, 
    commentstyle=\color{javapurple},
    showspaces=false,
    showtabs=false,
    xleftmargin=6.5pt,
    framexleftmargin=1.9pt,
    framexrightmargin=-1.8pt,
    framexbottommargin=2pt,
    framextopmargin=2pt,
    showstringspaces=false,
    escapeinside={<@}{@>}
}

\section{Embeddings for Code}\label{sec:emb}

An \emph{embedding} refers to a real-valued vector representation
for an input. An embedding function $E: \mathcal{X} \rightarrow \mathbb{R}^d$ takes
an input $x$ in the domain of $\mathcal{X}$ and produces its corresponding vector representation in a $d$-dimensional vector space.
This vector is said to be \emph{distributed}~\cite{bengio2013representation}, where
each dimension of the vector is not attributed to a specific
hand-coded feature of the input, but rather the ``meaning'' of the
input is captured by the vector as a whole.

Embeddings present multiple appealing properties. They are
more expressive than local representations, such as one hot encodings, as values along
each dimension are continuous~\cite{bengio2013representation}.
Embeddings can also be
learned, which makes them applicable to different domains where
we have example data. One possibility is to learn these embeddings using a
neural network, such that the function $E$ uses a network's learned
weights.

\subsection{Running Example}
We present a running example to illustrate some of the key concepts for the
use of embeddings in code search. Suppose we want to produce a vector
that can successfully represent the code snippet below.

\begin{lstlisting}[keywordstyle=\ttfamily]
for (entry : map.entrySet()) {
	System.out.println(entry);
}
\end{lstlisting}

One possible approach is to treat this source code as text, and \emph{tokenize}
this input into a collection of individual words. The extent of tokenization
(and filtering certain words) may depend on the specific model
design. For this example, we will tokenize based on standard English
conventions (e.g. white-space, punctuation) and punctuation relevant to code
(e.g. snake and camel case). The code snippet can now be treated as the
collection of words.

\begin{lstlisting}[keywordstyle=\ttfamily]
for entry map entry set system out println entry
\end{lstlisting}


One approach to learning token embeddings is with an unsupervised model.
One popular technique is \emph{word2vec}, which implements a skip-gram model~\cite{fasttext, mikolov2013distributed}. In the skip-gram
model, the embedding for a \emph{target} token is used to predict embeddings of \emph{context} tokens within a fixed window size.
In our example, given the embedding for the token \verb|set| and a window size
2, the skip-gram model would learn to predict the embeddings for the tokens
\verb|map|, \verb|entry|, \verb|system|, and \verb|out|. The objective of this
process is to learn an embedding matrix $T$, where each row corresponds to the
embedding for a token in the vocabulary, and where two embeddings are similar if
the corresponding tokens often occur in similar contexts.

At this point, we can map each word in our tokenized code snippet to
its corresponding embedding. For example,
\verb|for| may be represented by $\langle 0.2, -1.0, 3.8 \rangle$, and \verb|entry|
may be represented by $\langle 0.8, 0.9, -2.0 \rangle$.

\subsection{Bags and Sequences of Embeddings}
The next step in our procedure will be to combine the token-level embeddings for
our code snippet into a single embedding that appropriately represents the
snippet as a whole. We discuss two possible approaches to doing so using
standard neural network architectures.

So far we have not discussed the impact of the token order in our snippet.
We can decide to treat the words as a bag, occasionally called a multiset,
and ignore order. In such a case, our example

\begin{lstlisting}[keywordstyle=\ttfamily]
for entry map entry set system out println entry
\end{lstlisting}

would be equivalent to every other permutation, such as

\begin{lstlisting}[keywordstyle=\ttfamily]
entry for map println entry set out entry system
\end{lstlisting}

A corresponding bag-based neural network would compute the representation for
our code snippet without regard to token order. One simple example of such
a bag-of-words-based architecture is one where we use $T$, the matrix of learned
token embeddings, to look up the embedding for each word in the tokenized
example and then average (either simple or weighted) these vectors to obtain a single output vector.

In contrast, a neural network may instead consume the tokens in an input
as a sequence, such that
the ordering of elements is significant.
We provide details
on one common approach to handling sequence-based inputs: recurrent
neural networks (RNN)~\cite{goodfellow2016deep}.

An RNN starts with an initial hidden state, often initialized randomly,
represented as $h_0$, and processes the words in the input sequentially
one by one. In our example, the two permutations of the tokenized code snippet
are no longer equivalent.

After processing each word, the RNN updates the hidden state.  If
the $t^{\rm th}$ word in the sequence is $w_t$ and the hidden state after
processing the words before $w_t$ is $h_{t-1}$, then the next hidden state after
processing $w_t$ is obtained as follows:

\begin{align}
h_t = {\rm tanh}(W . [h_{t - 1};w_t ])
\end{align}

where $W$ is a matrix whose parameters are learned, $[x;y]$ is the vector
obtained by concatenating the vector $x$ and $y$, and
${\rm tanh}(x) = \frac{e^x - e^{-x}}{e^x + e^{-x}}$
is a non-linear activation function which ensures that
the value of $h_t$ lies between -1 and 1.

There are multiple approaches to obtain a snippet-level embedding using this
RNN. For example, one model might take the last hidden state $h_n$ as the
snippet embedding. Another could collect the hidden states $h_i$ and apply a reduction
operation such as dimension-wise max or mean to produce the snippet embedding.

The network described above
is a simple RNN; in practice, an RNN is implemented by using a more complex
function on $h_{t-1}$ and $w_t$. An example of such an RNN is
long-short term memory (LSTM)~\cite{hochreiter1997long}.

\subsection{Bi-Modal Embeddings}
So far, we have only discussed how to produce a representative vector given a code snippet.
However, neural code search uses embeddings for both code snippets and the
user's natural language query. This means that our embedding approach must be
able to represent both the code \verb|for (entry : map.entrySet) ... |, and
the query \textit{``how to iterate through a hashmap''}, which is expressed in
natural language.
Such embeddings that relate two different kinds of data are called bi-modal
embeddings~\cite{allamanis2015bimodal}.

The computation of bi-modal embeddings of a code snippet and its natural
language description can be abstractly formulated as two functions:
$E_c\colon \mathcal{C} \rightarrow \mathbb{R}^d$ and $E_q\colon \mathcal{Q} \rightarrow
\mathbb{R}^d$, where $\mathcal{C}$ is the domain of code snippets, $\mathcal{Q}$ is the domain of
natural language descriptions, $\mathbb{R}^d$ is a real-valued vector of length
$d$, $E_c$ is an embedding function that maps a code snippet to a $d$-dimensional
vector, and $E_q$ is an embedding function that maps a natural language
description to a vector in the same vector space. The goal is to learn the
functions $E_c$ and $E_q$ such that for some similarity measure ${\rm\it sim}$,
such as cosine similarity~\cite{nncosine},
${\rm\it sim}(E_c(c), E_q(q))$ is maximized for a code
snippet $c$ and its corresponding natural language description $q$.
Alternatively, for unsupervised models, such as \ncs{}, $E_c$ and $E_q$
may be instantiated with the same token-level embedding matrix $T$,
as shown in \Cref{fig:example}.

\subsection{Applying Embeddings to Code Search}
Given $E_c$ and $E_q$, code search can be performed given the user query and a code corpus.


\Cref{fig:example} illustrates how embeddings are used in code search.
The code embedding function $E_c$ is used to convert each candidate code snippet in
our search corpus into a vector.

For example, given the code snippet
\begin{lstlisting}
public void <@\textbf{\textcolor{javapurple}{forValuesAtSameKey}}@>(Map <K, V> map, ...) {
	...
}
\end{lstlisting}
in our search index,
an unsupervised $E_c$ (labeled \ncs{} in the figure) returns the
vector representation $\langle 0.94, -0.2, 1.2 \rangle$.
All the snippets in a corpus can be embedded in a similar fashion
and used to construct an index that allows for fast lookups based on a similarity metric.

The user query can be similarly embedded using $E_q$.
For example, \textit{``How do I iterate through a hashmap?''} is mapped
to the vector $\langle 0.2, 0.4, 0.5 \rangle$.
To retrieve relevant code snippets, the code embeddings index can be searched
based on similarity to the query embedding.
The top $N$ results based on this similarity are returned.

There are a number of possible neural architectures used to learn $E_q$ and
$E_c$, and we will explore several of them in this paper.

\section{Neural Code Search Models}\label{sec:models}
We now introduce each of the neural techniques explored in this paper.

\subsection{\ncs{}}\label{sec:ncs}
In \ncs{}~\cite{sachdev2018retrieval},
a specific technique named after the general concept of
\textit{Neural Code Search},
the embedding functions $E_c$ and $E_q$
are implemented using a combination of token-level embeddings using fastText~\cite{fasttext},
which is similar to \textit{word2vec}~\cite{mikolov2013distributed},
and conventional information retrieval techniques, such as TF--IDF~\cite{IR,tfidf}.
As such this technique does not use conventional deep
neural networks nor supervised training.  \ncs{} computes an embedding
matrix $T \in \mathbb{R}^{|V_c|\times d}$ , where $|V_c|$ is the size of the
code token vocabulary, $d$ is the chosen dimensionality of token embeddings,
and the $k^{\rm th}$ row in $T$ is the embedding
for the $k^{\rm th}$ word in $V_c$.
Once the matrix $T$ has been computed using fastText,
it is \emph{not} further modified using supervised training.

\ncs{} applies the \emph{same} embedding matrix $T$ to
\emph{both} the code snippets and the query as follows.
Let $c = \{c_1, \ldots, c_n\}$ and $q = \{q_1, \ldots, q_m\}$ represent the code snippet and
query, respectively, as a multiset (i.e. order insensitive) of tokens.
\ncs{} generates a bag of embedding
vectors $\{T \lbrack c_1 \rbrack, \ldots, T \lbrack c_n \rbrack \}$
for the code snippet and $\{T \lbrack q_1 \rbrack, \ldots, T \lbrack q_m \rbrack \}$
for the query, where $T \lbrack w \rbrack$ is the embedding vector
in the matrix $T$ for the token $w$.

To combine the bag of code token embeddings into a single code vector $e_c$, \ncs{}
sums the embeddings for the set of unique tokens weighed by their corresponding
TF--IDF weight. The TF--IDF weight is designed to increase the weight
of tokens that appear frequently in a code snippet, and decrease the weight of tokens that appear too frequently globally across all of the code corpus.
We elide the classical TF--IDF weighing formula here
for brevity.

\begin{figure}
	\centering
	\includegraphics[width=\columnwidth]{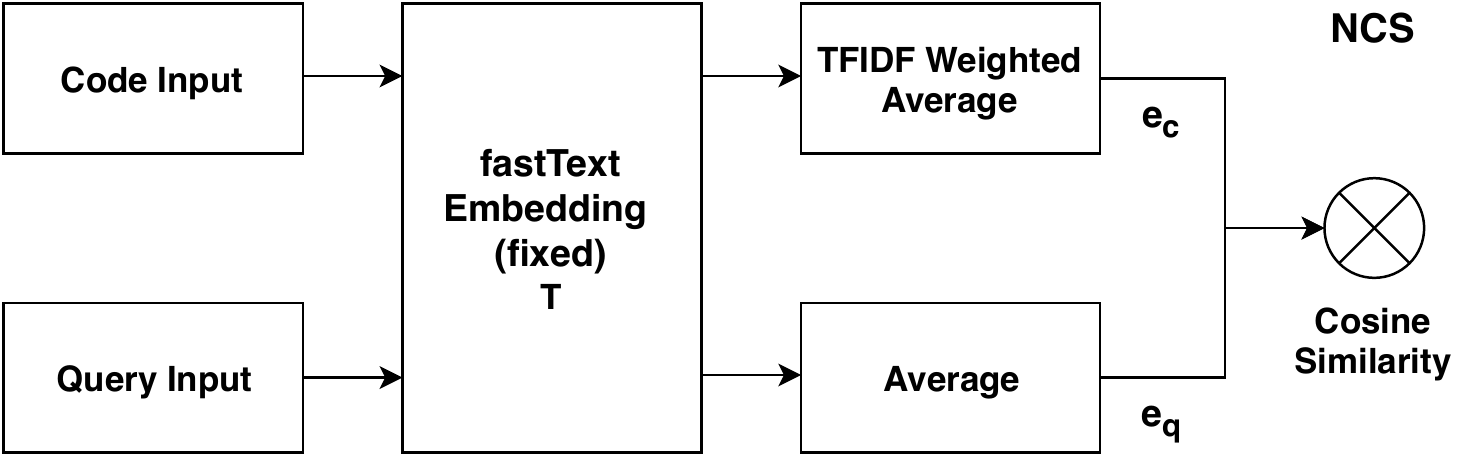}
	\caption{
		\ncs{} embeds the code and query input with the fastText~\cite{fasttext}
		embeddings. The code sentence embedding e\textsubscript{c} is computed from
		the bag of code embeddings with TF-IDF weights. The query sentence embedding
		e\textsubscript{q} is produced by averaging the bag of query embeddings.
	}
	\label{fig:ncs-diagram}
\end{figure}

For the query, \ncs{} averages the bag of query token embeddings
into a single query vector $e_q$.\footnote{
The authors of \ncs{}
also introduce a variant of their model that heuristically extends
user queries using code and natural language token co-occurrences.
We do not use this heuristic extension in order to directly observe
the impact of extending training with natural language supervision.
}
%
%
%
%
The high level architecture of the \ncs{} model is illustrated in
\Cref{fig:ncs-diagram}.
%

\subsection{\eu{}: A Supervised Extension of \ncs{}}
\label{sec:eu}
We will introduce \eu{} next, as it is a supervised minimal extension of the \ncs{} technique.
In this model, we use supervised learning to modify the initial
token embedding matrix $T$ and produce two embedding matrices, $T_c$
and $T_q$, for code and query tokens, respectively.
We also replace the TF-IDF weighing of
code token embeddings with a learned attention-based weighing scheme.
We refer to this extended approach as \emph{Embedding Unification} (\eu{}).

We assume that an aligned corpus of code snippets and their natural language
descriptions is available for training. We denote this corpus as a collection
of $(c, q)$, where $c$ is bag of tokens $c_1, \ldots, c_n$ from a code snippet and
$q$ is a bag of tokens from its corresponding natural
language description.

The functions $E_c$ and $E_q$ are constructed as follows.
Let $T_q \in \mathbb{R}^{|V_q|\times d}$ and $T_c \in \mathbb{R}^{|V_c|\times d}$
be two embedding matrices mapping each word from the natural language description
(specifically the docstrings and the query) and code tokens, respectively, to a
vector of length $d$. The two matrices are initialized using the same initial
weights, $T$, and modified separately during training.

We apply the respective embedding matrices to each element in the paired
corpus, such that for a code snippet $c$ we obtain a bag of embedding vectors
$\{T_c \lbrack c_1 \rbrack, \ldots, T_c \lbrack c_n \rbrack \}$,
and similarly for each description $q$. We compute a simple average
to combine the query token embeddings into a single vector. The simple averaging
is also present in \ncs{} and we found it to outperform attention-based weighing
during experiments.

To combine each bag of code token vectors into a single code vector that
captures the semantic meaning of the corresponding entity,
we use an attention mechanism~\cite{code2vec} to compute a weighted
average. The attention weights, $a_c \in \mathbb{R}^d$,
is a $d$-dimensional vector, which is
learned during training. $a_c$ acts as a learned counterpart
to the TF-IDF weights in \ncs{}.

Given a bag of code token embedding vectors $\{e_1, \ldots, e_n\}$, the attention
weight $\alpha_i$ for each $e_i$ is computed as follows:
\begin{eqnarray}
\alpha_i &=& \frac{\exp(a_c . e_i^\intercal)}{\sum_{i =1}^n \exp(a_c . e_i^\intercal)}
\end{eqnarray}

Here we compute the attention weight for each embedding vector as the softmax
over the inner product of the embedding and attention vectors.

The summary code vector of a bag of embedding vectors is then computed as the
sum of the embedding vectors weighted by the attention weights $\alpha_i$:
\begin{align}
e = \sum_{i=1}^n \alpha_i e_i
\end{align}

$e$ corresponds to the output of $E_c$.

\begin{figure}
	\centering
	\includegraphics[width=\columnwidth]{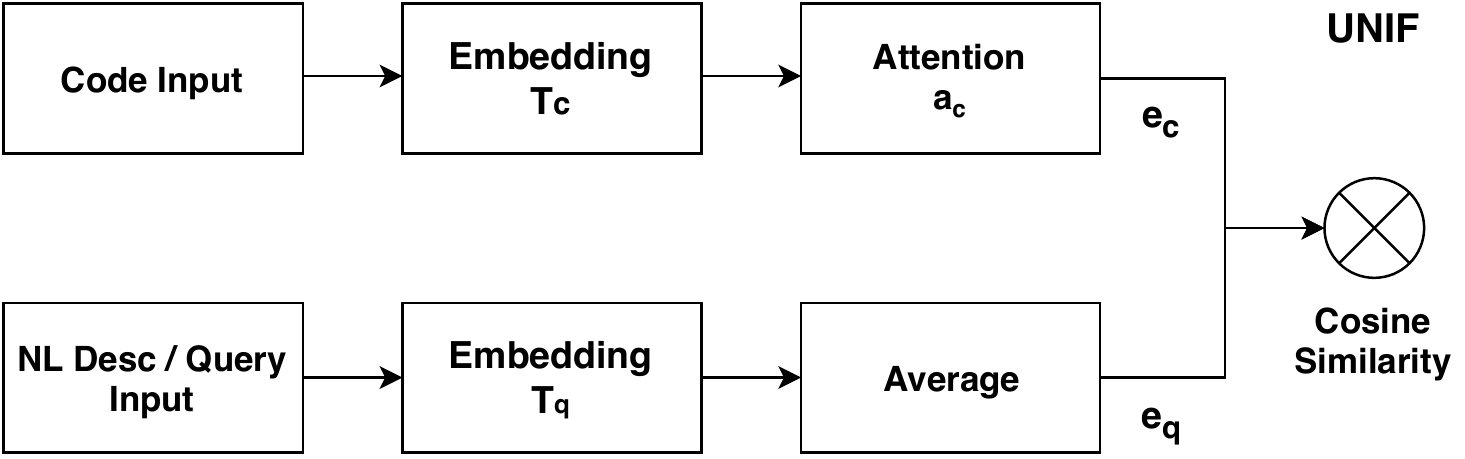}
	\caption{
		The UNIF network uses attention a\textsubscript{c}
		to combine per-token embeddings T\textsubscript{c} and produce the code sentence embedding e\textsubscript{c}. The query sentence embedding e\textsubscript{q} is produced by averaging the bag of query embeddings T\textsubscript{q}. Both T\textsubscript{c} and T\textsubscript{q} are initialized with the fastText~\cite{fasttext} embeddings and are further fine-tuned during training.
	}
	\label{fig:embedding-unification-diagram}
\end{figure}

Our training process learns parameters $T_q$, $T_c$, and $a_c$
using classic backpropagation.
\Cref{fig:embedding-unification-diagram} shows a high level diagram
of \eu{}.

\subsection{\codenn{}}
\label{sec:codenn}

Similar to \eu{}, \codenn{}~\cite{deepcodesearch} also models
both $E_c$ and $E_q$ using neural networks and employs supervised
learning; however, the networks used are more sophisticated
and deep. We adhere to the original authors' naming and
refer to this model as \codenn{},
short for \textit{Code Description Embedding Neural Network}.
Instead of treating a code snippet as a bag of tokens, \codenn{}
extracts a sequence of words from the name of the method containing the code
snippet, the sequence of API calls in the snippet, and a bag of tokens from the code snippet.
The word sequence from a method name is extracted by splitting the method name
on camel-case and snake-case.

The method name sequence and API sequence are given as input to two separate
bi-directional long-short term memory (bi-LSTM) networks~\cite{hochreiter1997long}.

After applying two separate LSTMs to the method name and API sequences,
\codenn{} obtains two sequences of hidden states.  \codenn{} summarizes each
such sequence of hidden states to obtain a single vector.  For summarization,
\codenn{} uses the max-pooling function.

Each token in the bag of code tokens is given individually as input to a feed
forward dense neural network and the output vectors are max-pooled. A final code
embedding is then obtained by concatenating these three vectors (two from the
LSTMs and one from the feed-forward network) and feeding them to a dense
neural network which produces a single summary vector $e_c$.  All the above networks
together implement the function $E_c$.

\begin{figure}
	\centering
	\includegraphics[width=\columnwidth]{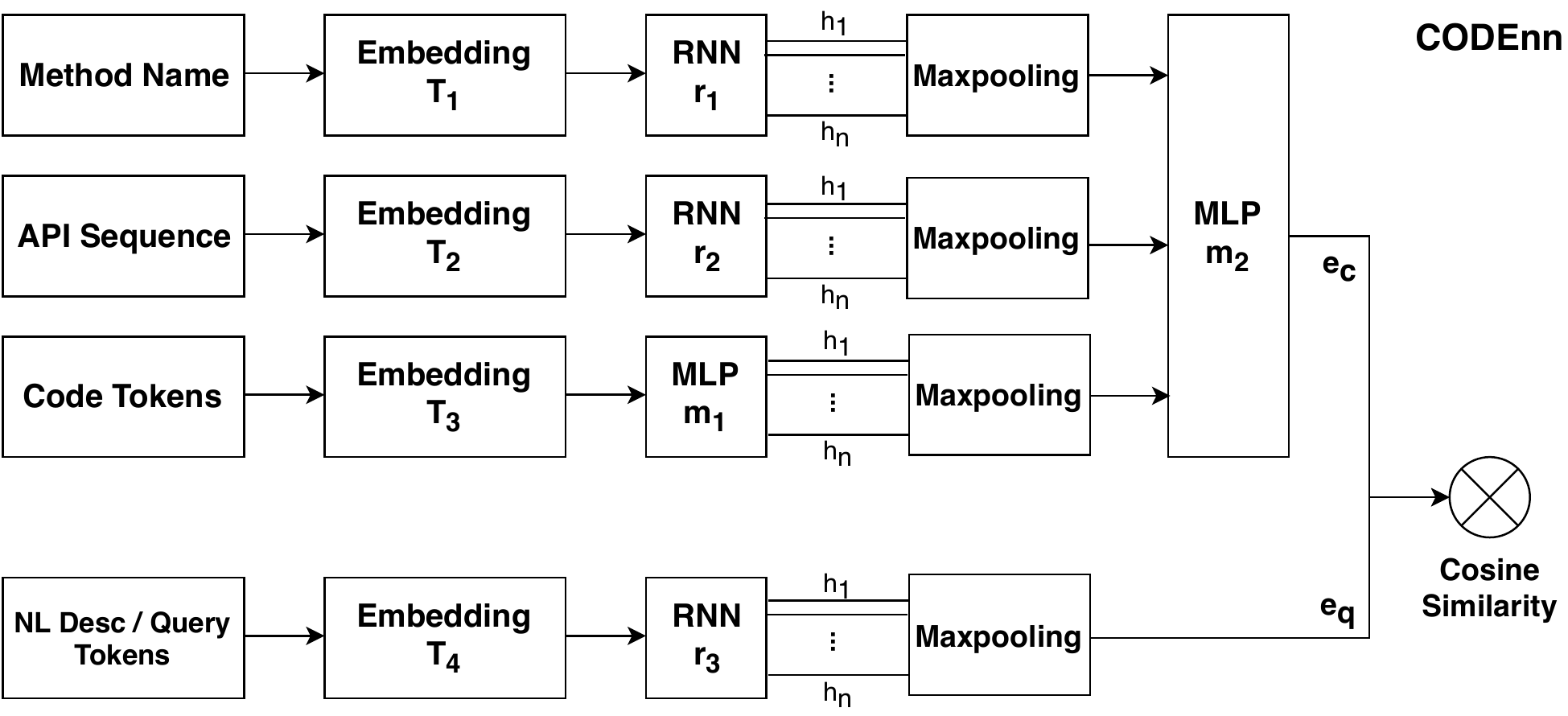}
	\caption{
		\codenn{}, the network proposed for code search uses
		RNNs to embed the method name (r\textsubscript{1}), API sequence (r\textsubscript{2}),
		and query (r\textsubscript{3}). It uses a feed forward network (MLP) to
		embed the code body tokens (m\textsubscript{1}), and combines this
		embedding with the method name and API embedding with another MLP
		(m\textsubscript{2}).
	}
	\label{fig:deep-code-search-diagram}
\end{figure}

\codenn{} implements the function $E_q$ using a bi-directional LSTM, which
takes as input sequence the description of the code snippet found in the
doc string to produce $e_q$. \Cref{fig:deep-code-search-diagram} provides an overview of the architecture.

\subsection{\githubmodel{}}
We introduce another supervised sequence-based deep neural network for code search,
described and implemented by the data science team at GitHub~\cite{githubblog}.
We will refer to this model as \githubmodel{}, short for \textit{Semantic Code Search}.

\githubmodel{} is divided into three separate training modules. A sequence-to-sequence gated
recurrent unit (GRU) network~\cite{cho2014learning} learns to generate a
docstring token sequence given a code token sequence. We refer
to this as the code-to-docstring model.

An LSTM network~\cite{merityRegOpt} learns a language model
for docstrings in the training corpus~\cite{sutskever2014sequence}. This model can be used to embed natural language and compute the probability of a given natural language input.



\begin{figure}
	\centering
	\includegraphics[width=\columnwidth]{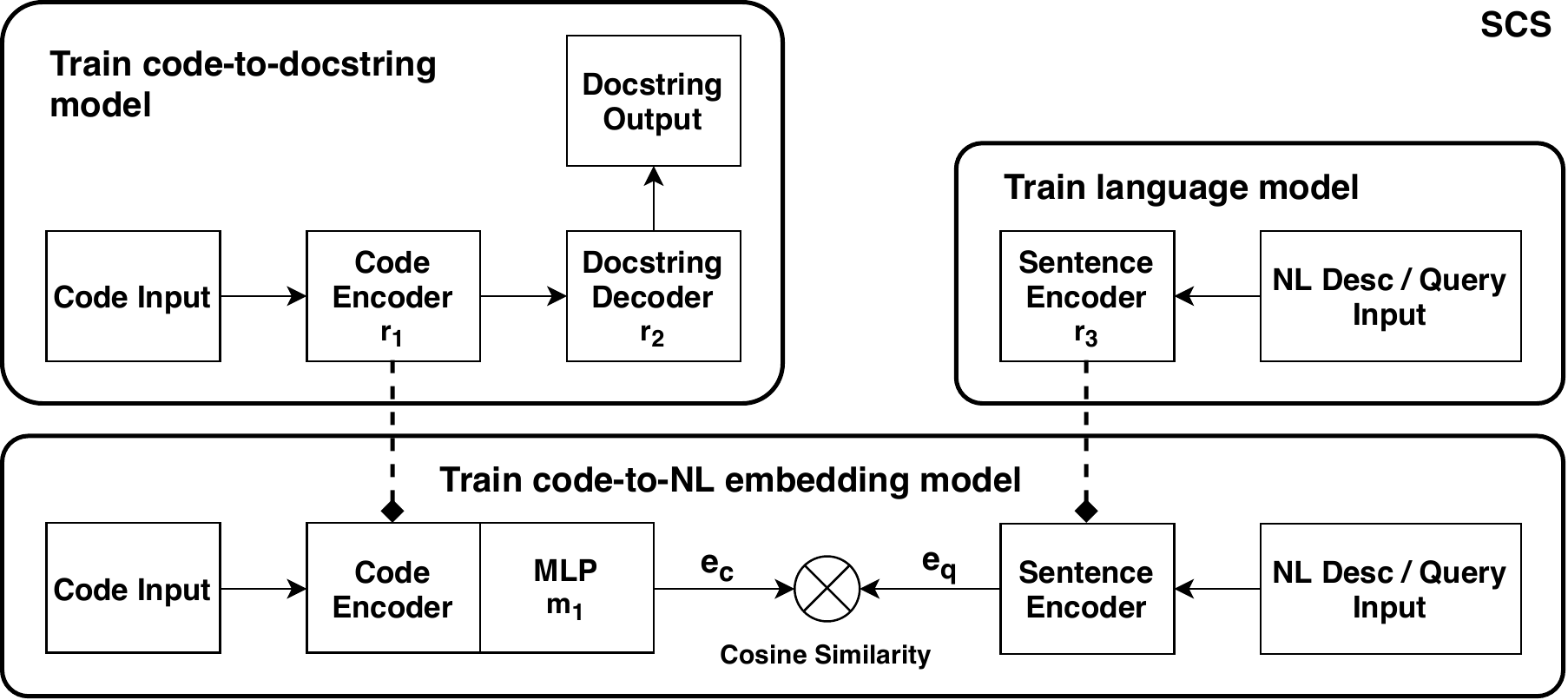}
	\caption{
		\githubmodel{} uses the encoder portion of the code-to-docstring sequence-based
		network to embed sequences of code tokens. Separately, it trains a language model to
		embed sequences of query tokens. A feed forward layer is added to the
		code encoder to transform code embeddings into query embeddings (derived
		from the language model).
	}
	\label{fig:gm-diagram}
\end{figure}

A final module learns a transformation (in
the form of a feed forward layer) to predict a query embedding given a sequence
of code tokens. To learn this transformation, the module takes the encoder portion of the
code-to-docstring model,
freezes its layers, and trains the
network on code sequence inputs and the corresponding query embedding produced
using the language model. A final training phase fine-tunes the network as a whole by
unfreezing the encoder layer for a few epochs. \githubmodel{} uses this fine-tuned encoder portion
of the code-to-docstring model as $E_c$ and the language model as $E_q$.
\Cref{fig:gm-diagram} provides an overview of the architecture.

\begin{table*}[!hbtp]
  \caption{
  Summary of details for models trained with supervision. In terms of
  network complexity (parameters and layers), from least
  to most complex, we have: \eu{}, \codenn{}, and \githubmodel{}.
  }
\centering
\def\arraystretch{1.25}
\small
\begin{tabular}{lll}
	\toprule
Model & Summary & Parameters \\
\hline

\eu{}
&
\begin{tabular}[c]{@{}l@{}}
Embeds code/query tokens. \\
Combines code embeddings with attention.
\end{tabular}
&
\begin{tabular}[c]{@{}l@{}}
	Embedding matrices $T_c$, $T_q$ (\Cref{fig:embedding-unification-diagram} T\textsubscript{c}, T\textsubscript{q}). \\
	Attention vector $a_c$ (\Cref{fig:embedding-unification-diagram} a\textsubscript{c}) \\
\end{tabular} \\
\hline

\codenn{}
&
\begin{tabular}[c]{@{}l@{}}
Embeds method name, API sequence and query as sequences. \\
Embeds a bag of tokens from body. \\
Combines method name, API, and token embeddings using another layer. \\
\end{tabular}
&
\begin{tabular}[c]{@{}l@{}}
	Embedding matrices $T_1$, $T_2$, $T_3$, $T_4$ (\Cref{fig:deep-code-search-diagram} T\textsubscript{1}, T\textsubscript{2}, T\textsubscript{3}, T\textsubscript{4})\\
	Bi-directional LSTM parameters for RNNs (\Cref{fig:deep-code-search-diagram} r\textsubscript{1}, r\textsubscript{2}, r\textsubscript{3}) \\
	MLP parameters (\Cref{fig:deep-code-search-diagram} m\textsubscript{1}, m\textsubscript{2})
\end{tabular} \\
\hline

\githubmodel{}
&
\begin{tabular}[c]{@{}l@{}}
	Embeds sequence of code tokens. \\
	Embeds sequence of query tokens. \\
	Transforms code embedding into query token space. \\
\end{tabular}
&
\begin{tabular}[c]{@{}l@{}}
		GRU parameters for RNNs (\Cref{fig:gm-diagram} r\textsubscript{1}, r\textsubscript{2}) \\
		LSTM parameters for RNN (\Cref{fig:gm-diagram} r\textsubscript{3}) \\
		MLP parameters (\Cref{fig:gm-diagram} m\textsubscript{1})
\end{tabular} \\
\bottomrule
\end{tabular}

\label{table:model-comparison}
\vspace{-4mm}
\end{table*}

\Cref{table:model-comparison} provides an overview of the network details for
models that employ supervision when learning their $E_c$ and $E_q$:
\eu{}, \codenn{}, and \githubmodel{}.

\section{Evaluation Methodology}\label{sec:methodology}
\label{sec:methodology}
Our evaluation uses different datasets and benchmarks. We use the following
terminology throughout for clarity:

\begin{itemize}[leftmargin=*]
	\item \textbf{\emph{training corpus}} refers to a dataset of paired code and
	natural language. An example for natural language could be the code fragment's corresponding
	docstring. A training corpus is used to \emph{train} the models and
	may contain duplicate code / natural language pairs. An unsupervised model,
	such as \ncs{}, uses only the
  code fragments from a dataset.

  \item \textbf{\emph{search corpus}} refers to a dataset of unique
  code entries\footnote{Dataset is deduplicated after tokenization.}.
  Entries are unique in order to avoid repetition
  in search results. We apply a trained model to a search corpus to \emph{search}
  for the top results for a given user query. This dataset is used during the evaluation of the models.

  \item \textbf{\emph{benchmark queries}} refers to a set of evaluation queries used
  to gauge the performance of trained models. Each query in a benchmark
  is accompanied by a gold-standard code fragment result, which we use to score results
  retrieved from a search corpus by a trained model.

\end{itemize}

\noindent
Our evaluation uses
three different training corpora, two search corpora,
and two sets of benchmark queries.

\subsection{Training Corpora}

We use three training corpora for our experiments.

\textbf{\dcstrain{}} is the dataset publicly released by the authors
of \cite{deepcodesearch}. This corpus consists of approximately
16 million pre-processed Java methods and their corresponding docstrings,
intended for training.
The dataset includes four types of inputs:
method name sequences, API sequences, a bag of method body tokens,
and docstring sequences. We additionally derive another input by concatenating
the method name sequences to the API sequences and treating this concatenated
sequence as a bag of tokens. This derived input is used to train \eu{} and
\githubmodel{}.\footnote{Supervised models trained on \dcstrain{} train for 50 hours
on an Nvidia Tesla M40 GPU.}\footnote{This data has been generously made
available by the \codenn{} authors at
\url{https://drive.google.com/drive/folders/1GZYLT_lzhlVczXjD6dgwVUvDDPHMB6L7}
}

\textbf{\androidtrain{}} is an Android-specific corpus that we built by
collecting methods from approximately 26,109 GitHub repositories
with the Android tag. We took
all methods with an accompanying docstring (approximately 787,000 in total)
and used these as training data. Similar to \dcstrain{}, we
derive the four types of input collections (method name sequences, API sequences,
bag of method body tokens, and docstring sequences) necessary to train
\codenn{}. We train \eu{} and \githubmodel{} on the input sequence
generated for \ncs{}, which uses a parser to qualify method names with their
corresponding class, method invocations, enums, string literals
and source code comments, while ignoring variable names, and applies
a camel and snake case tokenization~\cite{sachdev2018retrieval}.\footnote{Supervised
models trained on \androidtrain{} train for 3 hours on an Nvidia Tesla M40 GPU.}

\textbf{\sotrain{}} is an Android-specific training
corpus that we built by collecting
Stack Overflow question titles and code snippets answers.
We prepared this dataset by extracting all Stack Overflow
posts with an Android tag from a data dump publicly released by Stack
Exchange~\cite{stackexchange}. The dataset is filtered on the following
heuristics: (1) The code snippet must not contain XML tags; (2) The code snippet
must contain a left parenthesis `(' to indicate presence of a method call; and
(3) The post title must not contain keywords like ``gradle'', ``studio'' and
``emulator''. After filtering, we end up with 451k Stack Overflow
title and code snippet pairs. This dataset is \emph{disjoint} from
the \ncsquestions{} benchmark queries described later on
in \Cref{sec:benchmark-queries}.

The goal of \sotrain{} is to serve as an alternate training corpus that is ideal
for our evaluation, which leverages Stack Overflow titles and
code snippets as benchmark queries and answers, respectively. By training
on this corpus, we can measure the potential for improvement
compared to training on a typical aligned corpus, which uses docstrings
as natural language.\footnote{Supervised
models trained on \sotrain{} train for 3 hours on an Nvidia Tesla M40 GPU.}

\subsection{Search Corpora}
We use two search corpora during our evaluation.

\textbf{\dcssearch{}}: 4 million \textit{unique} Java methods released by the
authors of \codenn{}.

\textbf{\androidsearch{}}: 5.5 million \textit{unique} Android methods
collected from GitHub.
This corpus is derived from the same 26,109 repositories used
to construct \androidtrain{}, but also includes methods that
do not have a docstring available.

\subsection{Benchmark Queries}\label{sec:benchmark-queries}
We use two sets of queries as
evaluation benchmarks for our models.
In both benchmark sets, the queries correspond to Stack Overflow titles and the
ground truth answers for each query are the accepted answer or highly voted
answer for the corresponding post, which we independently collected. This approach to
collecting ground truth answers was borrowed from the original \ncs{}
paper~\cite{sachdev2018retrieval}. The use of Stack Overflow titles as queries,
rather than a small set of initial keywords, aims to evaluate the
extent to which the different techniques successfully map natural language fragments
and code to a shared space, and was chosen to directly compare to
prior work~\cite{sachdev2018retrieval, deepcodesearch}.

\textbf{\dcsquestions{}} is a set of 50 queries used to evaluate
\codenn{} in the original paper. These queries correspond to
Stack Overflow titles for the top 50 voted Java programming questions.
The authors included questions that had a ``concrete answer'' in Java,
included an accepted answer in the thread with code, and were not
duplicate questions. When evaluating on this benchmark, models are trained
on \dcstrain{}.

\textbf{\ncsquestions{}} is a set of 287 Android-specific
queries used to evaluate \ncs{} in the original paper.
These questions were chosen by a script with the following criteria: (1)
the question title includes ``Android'' and ``Java'' tags; (2) there exists
an upvoted code answer; and (3) the ground truth code snippet has at least one
match in a corpus of GitHub Android repos. When evaluating on this
benchmark, models are trained on \androidtrain{}, unless otherwise specified.

\Cref{table:data-used} provides a summary of the training corpora, search
corpora, and benchmark queries used in our evaluation, and what
combinations we use for our results.

\begin{table*}[]
	\caption{
	Summary of data used.
	When evaluating on \ncsquestions{}, we use
	\androidsearch{} as search corpus and train on \androidtrain{} or
	\sotrain{}. When evaluating on \dcsquestions{}, we use \dcssearch{}
	as search corpus and train on \dcstrain{}.
	}
\begin{subtable}{2\columnwidth}
\subcaption{Training corpora}
\centering
\begin{tabular}{ccc}
\toprule
Dataset       & Code/Natural Language & Number of Observations\\ \midrule
\dcstrain{}~\cite{deepcodesearch}   & Method/docstring   & 16mm               \\
\androidtrain & Method/docstring       & 787k            \\
\sotrain      & Forum code snippet/Forum title      & 451k             \\ \bottomrule
\end{tabular}
\end{subtable}

\begin{subtable}{\columnwidth}
\subcaption{Search corpora}
\centering
\begin{tabular}{cc}
\toprule
Dataset       & Number of Entries\\ \midrule
\dcssearch{}~\cite{deepcodesearch}   &  4mm               \\
\androidsearch{} & 5.5mm       \\ \bottomrule
\end{tabular}
\end{subtable}
~
\begin{subtable}{\columnwidth}
\subcaption{Benchmark Queries}
\centering
\begin{tabular}{cc}
\toprule
Benchmark Queries & Number of Queries \\ \midrule
\dcsquestions{}~\cite{deepcodesearch}       & 50             \\
\ncsquestions{} & 287            \\ \bottomrule
\end{tabular}
\end{subtable}
\label{table:data-used}
\end{table*}

\subsection{Evaluation Pipeline}

\begin{figure}
	\centering
	\includegraphics[width=\columnwidth]{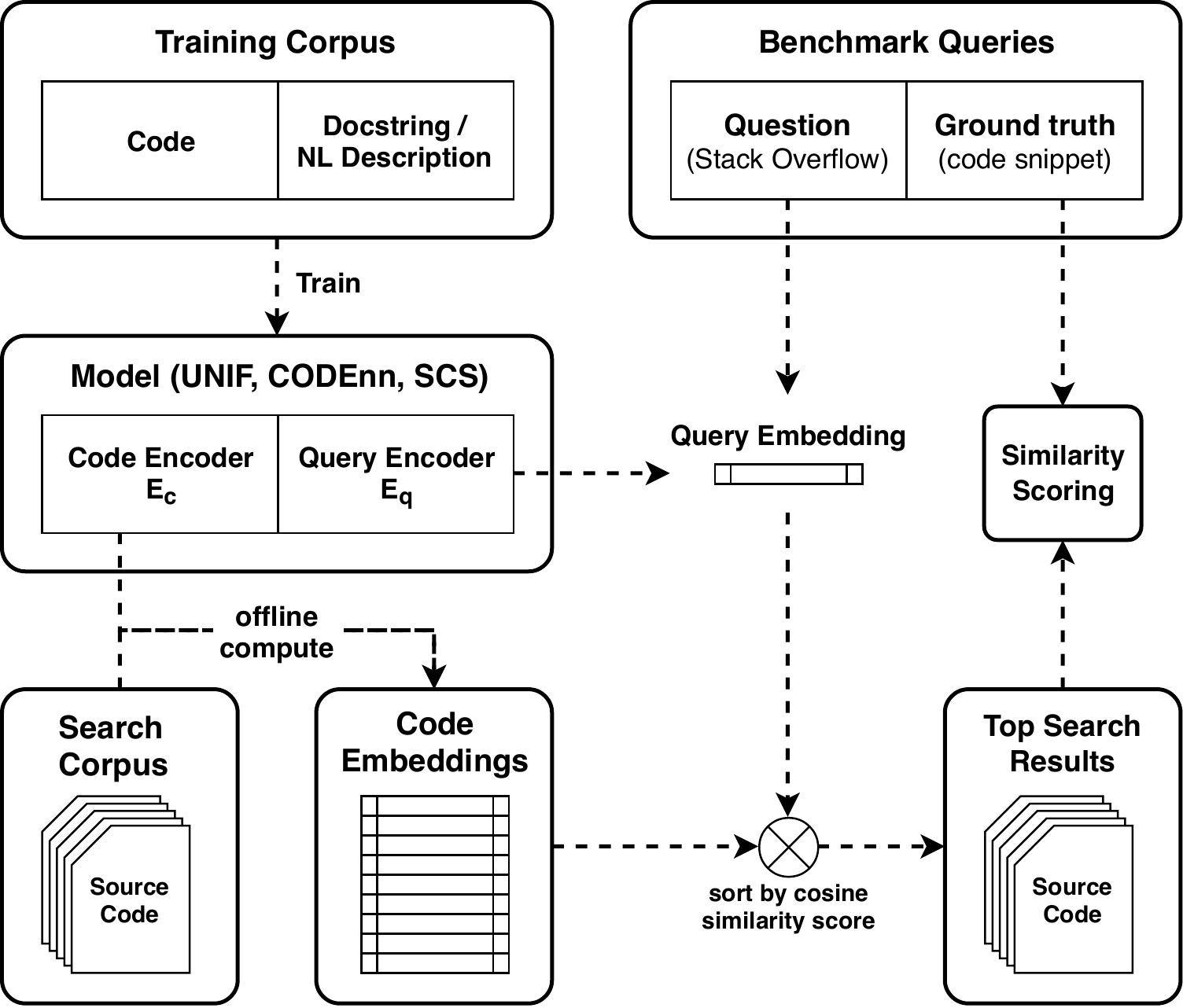}
	\caption{Evaluation pipeline.}
	\label{fig:evaluation_pipeline}
\end{figure}

We found that manually assessing the correctness of search results can be difficult
to do in a reproducible fashion, as deciding the relevance or correctness of a
code snippet relative to the input query can vary across authors and
people trying to reproduce our results.
As such, we decided to carry out our evaluation using an automated evaluation
pipeline. Our pipeline employs a similarity metric~\cite{aromaarxiv} between
search results and a ground truth code snippet to assess whether a query was
correctly answered. With this pipeline, we can scale our experiments to a much
larger set of questions, such as \ncsquestions{}, and assess correctness of
results in a reproducible fashion. We use code answers found on Stack Overflow
to provide a consistent ground truth for evaluation.\footnote{ We tried to
obtain code snippets marked as relevant from the original \codenn{} authors for
completeness, but they were unable to share them~\cite{gupersonalcommunication}.
} This approach was introduced by the authors of
~\cite{sachdev2018retrieval}. Figure \ref{fig:evaluation_pipeline} gives an
overview of this evaluation pipeline.

The automated pipeline does require that we pick a similarity threshold
to decide whether a query has been answered. To decide this value,
two authors manually assessed the relevance of the top 10 search results
for \dcsquestions{} produced by \codenn{} and \eu{}. This assessment was done
individually and conflicting decisions were cross-checked and re-assessed. Once
a final set of relevant results was determined, we computed the similarity
metric for each result with respect to the appropriate ground truth answer. This
yielded a distribution of scores that was approximately normal. We took the mean
and use this as the similarity threshold in our evaluation. This threshold
chosen produces evaluation metrics for \codenn{} that generally correspond to
those in its original paper~\cite{deepcodesearch}.


In our evaluation, we present the number of questions answered in the top
$k$ results. We consider the top 1, 5 and 10 results, and display the corresponding
number of questions answered as \emph{Answered@{1,5,10}}, respectively.




\section{Results}\label{sec:results}
We now present our study's results and the answer for each of the
research questions posed.

\subsection{RQ1}
\noindent{\emph{Does extending an effective
unsupervised code search technique with supervision based on
a corpus of paired code and natural language improve performance?}}

As detailed in \Cref{sec:eu}, \eu{} is an extension of \ncs{} that
adds supervision to modify embeddings during training and replaces the
TF-IDF weights used to combine code token embeddings with learned attention.
\Cref{table:rq1} shows that \eu{} answers more questions
across the board for \dcsquestions{}.
\eu{} improves the number of answers in the top 10 results for
\ncsquestions{} but performs slightly worse for answers in the top 1.
 We conclude that extending
a \ncs{}, an unsupervised technique, with supervision improves
performance for code search, but not uniformly across our datasets.

\begin{table}
  \caption{
  Number of queries answered in \dcsquestions{}
  in the top 1, 5, and 10 results improves when we extend
  \ncs{} (unsupervised) to \eu{} (supervised).
  For \ncsquestions{}, supervision increased results in the top 5 and top 10.
  }\label{table:rq1}

\centering
\footnotesize
\begin{tabular}{@{}ccccc@{}}
\toprule
Benchmark queries & Model        & Answered@1 & Answered@5 & Answered@10 \\ \midrule
\multirow{2}{*}{\dcsquestions{}} & \ncs{}   & 15    & 29 & 37   \\
                                 & \eu{}  & \textbf{22}    & \textbf{39} & \textbf{43}   \\ \midrule
\multirow{2}{*}{\ncsquestions{}} & \ncs{}   & \textbf{33}    & \textbf{74} & 98   \\
																 & \eu{}  & 25    & \textbf{74} & \textbf{110}   \\ \bottomrule
\end{tabular}
\end{table}

%
%

\subsection{RQ2}
\noindent{\emph{Do more sophisticated networks
improve performance in supervised neural code search?}}

When selecting possible supervised techniques, neural code search system designers
may choose to incorporate more sophisticated architectures such
as \codenn{} or \githubmodel{}, or favor a simple architecture, such as that used in \eu{}.
In order to navigate this question, we consider the number of queries
answered by different techniques, as well as their computational cost.

We first compare model performance in terms of the number of queries
correctly answered. \Cref{table:rq2-perf} shows that
\eu{}, which uses a simple bag-of-words approach, outperformed both
\codenn{} and \githubmodel{} on both benchmark query sets. \codenn{} performed
better than \githubmodel{} in both cases.

\begin{table}
	\caption{The evaluation results on both benchmarks show that
	 \eu{} outperforms more sophisticated sequence-based networks
	 such as \codenn{} and \githubmodel{}.
	}\label{table:rq2-perf}

\centering
\footnotesize
\begin{tabular}{@{}ccccc@{}}
\toprule
Benchmark queries & Model        & Answered@1 & Answered@5 & Answered@10 \\ \midrule
\multirow{3}{*}{\dcsquestions{}} & \eu{}   & \textbf{22}    & \textbf{39}    & \textbf{43}     \\
                                 & \codenn{} & 16    & 31    & 39     \\
																 & \githubmodel{}     & 9     & 17    & 21     \\ \midrule
\multirow{3}{*}{\ncsquestions{}} & \eu{}   & \textbf{25} & \textbf{74} & \textbf{110}    \\
																 & \codenn{} & 22 & 60 & 82    \\
																 & \githubmodel{}     & 9 & 19 & 34    \\ \bottomrule
\end{tabular}
\end{table}

%
%
%

Fully explaining the performance differential in neural systems is an open
research question~\cite{melis2018towards, zhang2018interpretable,
thomas2018interpretable}. As such, providing an in-depth discussion of the
factors influencing \eu{}'s out-performance is out-of-scope. However, we
conjecture that \codenn{}'s and \githubmodel{}'s use
of token order leads to overfitting, while \eu{}'s bag-oriented embeddings
results in a form of regularization that generalizes better during evaluation.
However, we note that this conjecture is not something we have explored
with experiments.

A second consideration in the complexity tradeoff is the cost of
computation across architectures. More sophisticated networks,
in particular those that consume sequences and thus maintain intermediate
state, are typically slower due to the amount of computation they perform.
A simpler architecture can provide faster inference and reduced training time.

As a way to quantify the increase in computation,
we measured the time to embed a sample of
code and natural language descriptions from \dcstrain{}. Code entries
were embedded in a batch of size 1000, while queries were embedded one at a time,
to reflect the expected behavior in a code search system, where code is embedded
offline and stored in an index and queries are embedded in real-time.

\Cref{table:rq2-inference-time} shows
the ratios of inference times \textbf{relative to \eu{}}
in each column, such that a value above 1.0 indicates slower inference for that
column. Sequence-based networks
such as \codenn{} and \githubmodel{} take longer to embed both code and natural
language inputs.

Note that all systems can embed code fragments offline, and the amount of time
to embed a query in real-time is relatively small. However, the relative increases
in time to embed code and query when going from the simple \eu{} to
more complex networks (\codenn{} and \githubmodel{}) highlight the increased
amount of computation that those networks perform.

\begin{table}[!h]
\caption{
\emph{Time ratios relative to \eu{}} to embed sampled code and natural language
from \dcstrain{}.
Values above 1 represent an inference time longer than \eu{}'s for
that column.
}\label{table:rq2-inference-time}
\small
\begin{tabular}{@{}ccccc@{}}
\toprule
Model  & code (CPU) & code (GPU) & query (CPU) & query (GPU) \\ \midrule
\codenn{} & 11.72 & 58.55 & 103.83 & 10.75 \\
\githubmodel{}     & 11.92 & 35.01 & 105.10 & 15.94     \\ \bottomrule
\end{tabular}
\end{table}

\subsection{RQ3}
\noindent{\emph{How effective is supervision based on docstrings
as the natural language component of the training corpus?}}

The supervised techniques presented so far use the same kind of natural
language during training: docstrings. Docstrings are used as a proxy
for user queries and allow neural code search practitioners to collect
sizeable aligned datasets for training. However, \Cref{tab:results} shows that when training on \androidtrain{}, search performance did not always improve, contrary
to expectations.

In this experiment, we use an alternate idealized training corpus, \sotrain{}, which
is drawn from the same source as our benchmarks \ncsquestions{},
but is still disjoint from the queries. Intuitively, the performance
attained with this corpus provides a measure for how much supervised
techniques could improve search, given a training corpus that matches
eventual user queries.

\Cref{table:rq3} shows that when we train on \sotrain{}, all supervised
techniques improve significantly (with one exception, queries answered in the
top 1 using the \githubmodel{}). This highlights the impressive search
performance that a supervised technique could deliver, if given
access to an ideal training corpus with a natural language component
that better matches user queries.

\begin{table}[!h]
\centering
\caption{
The number of \ncsquestions{} answered in the top 1, 5, and 10 when the
supervised techniques were trained on our idealized \sotrain{} corpus.
Search performance increases substantially, demonstrating the potential
for supervised techniques, when given access to a training corpus
that resembles eventual user queries.
}\label{table:rq3}
\begin{tabular}{@{}cccc@{}}
\toprule
Model  & Answered@1 & Answered@5 & Answered@10 \\ \midrule
\eu{}   & 25 $\shortrightarrow$ \bf{104} & 74 $\shortrightarrow$ \bf{164} & 110 $\shortrightarrow$ \bf{188}      \\
\codenn{} & 22 $\shortrightarrow$ 36        & 60 $\shortrightarrow$ 91 & 82  $\shortrightarrow$ 117              \\
\githubmodel{}     & 9 $\shortrightarrow$ 11 & 19 $\shortrightarrow$ 24 &     34 $\shortrightarrow$ 47         \\ \bottomrule
\end{tabular}
\end{table}

\Cref{tab:results} provides a comprehensive performance summary detailed previously
in each research question.

\begin{table*}
\begin{center}
\def\arraystretch{1.20}
\small
\caption{
Summary of evaluation results.}
\begin{tabular}{c|c|c|c|c|c|c|c}
\toprule
Benchmark queries                & Search corpus                 & Training corpus                  & Model          & Answered@1 & Answered@5 & Answered@10 & MRR \\
\midrule
\multirow{4}{*}{\dcsquestions{}} & \multirow{4}{*}{\dcssearch{}} & Unsupervised                     & \ncs{}         & 15 & 29 & 37 & 0.437 \\
\cline{3-8}
                                 &                               & \multirow{3}{*}{\dcstrain{}}     & \eu{}          & \bf{22} & \bf{39} & \bf{43} & \bf{0.582} \\
\cline{4-8}
                                 &                               &                                  & \codenn{}      & 16 & 31 & 39 & 0.456 \\
\cline{4-8}
                                 &                               &                                  & \githubmodel{} & 9 & 17 & 21 & 0.166 \\
\midrule
\multirow{7}{*}{\ncsquestions{}} & \multirow{7}{*}{\androidsearch{}} & Unsupervised                     & \ncs{}         & 33 & 74 & 98 & 0.189 \\
\cline{3-8}
                                 &                                   & \multirow{3}{*}{\androidtrain{}} & \eu{}          & 25 & 74 & 110 & 0.178 \\
\cline{4-8}
                                 &                                   &                                  & \codenn{}      & 22 & 60 & 82 & 0.150 \\
\cline{4-8}
                                 &                                   &                                  & \githubmodel{} & 9 & 19 & 34 & 0.124 \\
\cline{3-8}
                                 &                                   & \multirow{3}{*}{\sotrain{}}      & \eu{}          & \bf{104} & \bf{164} & \bf{188} & \bf{0.465} \\
\cline{4-8}
                                 &                                   &                                  & \codenn{}      & 36 & 91 & 117 & 0.215 \\
\cline{4-8}
                                 &                                   &                                  & \githubmodel{} & 11 & 24 & 47 & 0.138 \\
\bottomrule
\end{tabular}
\end{center}
\label{tab:results}
\end{table*}

\section{Threats to Validity}\label{sec:threats}

Our evaluation shows that a supervised extension of \ncs{}
performed better than the original unsupervised version.
There may exist other unsupervised techniques
which require more in-depth modification to successfully take
advantage of supervision. Our goal, however, is not to show that
our minimal extension is guaranteed to improve any unsupervised technique,
but rather that it improves \ncs{} specifically.

\eu{} is presented as a simple alternative to state-of-the-art models.
We explored two techniques from current literature and show that
\eu{} outperforms them. More sophisticated architectures may successfully
outperform \eu{} but we believe that our result highlight the importance
of exploring parsimonious configurations first.


We relied on an automated evaluation pipeline to provide reproducible
and scalable evaluation of code search results. With this we scaled
evaluation to a much larger set of benchmark queries (\ncsquestions{}).
While performance may vary depending on the similarity threshold
and algorithm chosen,
we derived our similarity threshold choice through manual evaluation
of answers produced by two techniques (\codenn{} and \eu{}) and
believe it correlates well with human judgment. This threshold and
the similarity algorithm used with it
produces evaluation results for \codenn{} that roughly correspond
to those found in its original paper~\cite{deepcodesearch}.

\section{Related Work}\label{sec:relatedwork}
Recent works from both academia and industry have explored the realm of code
search. \ncs{}~\cite{sachdev2018retrieval} presented a simple yet effective
unsupervised model. \codenn{}~\cite{deepcodesearch} and
\githubmodel{}~\cite{githubblog} provided a deep learning approach by using
sophisticated neural networks. These systems build on the idea
of bimodal embeddings of source code and natural language.
Sachdev et al~\cite{sachdev2018retrieval} also compared neural code search to
traditional IR techniques such as BM25 and showed that neural techniques
outperform when results are post-ranked (a common technique in
IR).

Existing work in natural language processing
~\cite{artetxe2017learning,conneau2017word,grave2018unsupervised} has explored
constructing embeddings for two languages with little bilingual data.
It is possible that some of these techniques might be applicable to
the code search task, and address some of the issues we identified
during our analysis. However, in the code search task the embedding alignment
we care about is not just at the word (i.e. token) level, but rather
should aggregate successfully to whole code snippets and queries.

Other than code search, a line of work has explored enhancing developer productivity by exploiting an aligned corpus of code and natural language. Allamanis et al. ~\cite{allamanis2015bimodal} proposes a probabilistic model to synthesize a program snippet from a natural language query. Bayou ~\cite{bayou} is a system that uses deep neural networks to synthesize code programs from a few API calls and a natural language input. CODE-NN ~\cite{iyer2016summarizing} uses LSTM networks to produce natural language descriptions given a code snippet.

Interest in applications of neural networks to software engineering has
increased significantly. Existing work has introduced neural networks to
identify software defects~\cite{dam2018deep}, guide program
synthesis~\cite{wang2018execution, balog2016deepcoder}, enable new
representations for program analysis~\cite{code2vec,allamanis2017learning},
facilitate code reuse~\cite{guo2017semantically}, and
automate code changes~\cite{tufano2019learning}. The models we present here
make use of these technologies to varying degrees to explore the design space
and impact of these choices on code search quality.

Other areas of software engineering have used neural networks to improve
performance in tasks that can be formulated as a search given an input
``query''.  For example, bug localization uses an input query (e.g. a bug
report, regression test outputs) to identify source files in a project's tree
that are relevant to the bug by using deep neural networks to create representations of
the input query and source tree files ~\cite{lam2017bug, wong2011effective, huo2016learning}.
Our research is complementary to this work as we evaluate the performance of different neural
network architectures on a search task. However, in contrast,
the code search tasks we target and evaluate span multiple source code
projects and rely exclusively on a natural language query as an input for search.


\section{Conclusion}\label{sec:conclusion}
In this paper we explored some of the design points made in previous works (e.g.
sequence-based models, docstring supervision).
We compared three state-of-the-art techniques
for neural code search with a novel extension of our own, and provided quantitative
evidence for key design considerations.

We showed that supervision,
shown by extending \ncs{} to \eu{}, can improve performance over an
unsupervised technique. We suggest baselining against an unsupervised neural code
search system and comparing incremental improvements, which should be weighed
against the time and resources required to collect
supervision data.

We found that \eu{} outperformed the more sophisticated
\codenn{} and \githubmodel{} models
on our benchmarks.
With this observation in mind, we suggest
evaluating simple architectures before incorporating
more sophisticated components such as RNNs into code search systems.

Finally, we showed that an ideal training corpus that resembled eventual
user queries provided impressive improvements for all supervised techniques.
We suggest considering the extent to which
a training corpus resembles eventual user queries
for optimal results, and exploring the possibility of better training
corpora, rather than assuming a code/docstring corpus will provide
the best performance.

\section*{Acknowledgements}
We would like to thank the authors of \codenn{}
for making their system and data public. Similarly,
we thank the authors of \githubmodel{} for making their blog post
and code available. We thank Celeste Barnaby,
J{\"u}rgen Cito and Vijay Murali for feedback on
an earlier draft of this paper.

\balance
\bibliographystyle{plainurl}
\bibliography{references}

\end{document}